\documentclass{jpp}
\usepackage{graphicx,natbib}

\ifCUPmtlplainloaded \else
  \checkfont{eurm10}
  \iffontfound
    \IfFileExists{upmath.sty}
      {\typeout{^^JFound AMS Euler Roman fonts on the system,
                   using the 'upmath' package.^^J}%
       \usepackage{upmath}}
      {\typeout{^^JFound AMS Euler Roman fonts on the system, but you
                   dont seem to have the}%
       \typeout{'upmath' package installed. JPP.cls can take advantage
                 of these fonts, if you use 'upmath' package.^^J}%
       \providecommand\upi{\pi}%
      }
  \else
    \providecommand\upi{\pi}%
  \fi
\fi

\usepackage{amsmath,amssymb}

\begin{document}
\title[Proton instabilities in weakly collisional plasmas]{Proton temperature-anisotropy-driven instabilities in weakly collisional plasmas:
Hybrid simulations}

\author[Hellinger and Tr\'avn\'\i\v cek]{Petr Hellinger\textsuperscript{1,2} and
Pavel M. Tr\'avn\'\i\v cek\textsuperscript{3,1,2}}

\affiliation{\textsuperscript{1}Astronomical Institute AS CR,
Bocni II/1401, CZ-14131 Prague, Czech Republic\\[\affilskip]
\textsuperscript{2} Institute of Atmospheric Physics, AS CR,
Bocni II/1401, CZ-14131 Prague, Czech Republic\\[\affilskip]
\textsuperscript{3} Space Sciences Laboratory, UCB, Berkeley, USA.}

\pubyear{2014}
\volume{?}
\pagerange{??}
\date{?; revised ?; accepted ?.}

\maketitle

\begin{abstract}
Kinetic instabilities in weakly collisional, high beta plasmas are investigated
using two-dimensional hybrid expanding box simulations with Coulomb collisions modeled
through the Langevin equation (corresponding to the Fokker-Planck one).
The expansion drives a parallel or perpendicular temperature anisotropy (depending on the orientation 
of the ambient magnetic field). For the chosen parameters the Coulomb collisions are important
with respect to the driver but are not 
strong enough to keep the system stable with respect to instabilities driven by the proton temperature 
anisotropy. In the case of  the
parallel temperature anisotropy the dominant oblique fire hose
instability efficiently reduces the anisotropy in a quasilinear
manner. In the case of the perpendicular temperature  anisotropy the dominant mirror instability generates
coherent compressive structures which scatter protons and reduce the temperature
anisotropy. For both the cases the instabilities generate temporarily enough wave energy 
so that the corresponding (anomalous) transport coefficients dominate over
the collisional ones and their properties are similar to those in
 collisionless plasmas.
\end{abstract}

\section{Introduction}

Large scale (turbulent) motion of the plasma often leads to changes in the magnitude
of the magnetic field (and possibly also in the particle densities). Such changes tend to generate
particle temperature anisotropies. In collisionless plasmas where no 
important wave activity of heat fluxes are present the two first adiabatic
invariants are expected to be conserved and, consequently, the parallel and perpendicular
temperatures (with respect to the ambient magnetic field) behave differently \citep{chewal56}.
 Coulomb collisions naturally reduce the temperature anisotropies
but may not be sufficient to keep the plasma in thermal equilibrium. 
The plasma system may eventually become unstable with respect to
temperature-anisotropy driven instabilities.
For example,
in the case of the almost collisionless solar wind plasmas
protons exhibit relatively large temperature anisotropies 
with signatures of bounds imposed by kinetic instabilities
\citep{hellal06,wickal13,baleal09,hetr14}.
Similar effects are expected in other astrophysical situations such as in 
the intercluster high-beta plasmas  
\citep{rosial11,santal14}
where Coulomb collision are expected to play important role but
where even a very weak temperature anisotropy may drive the system unstable.

Coulomb collisions and kinetic instabilities are relatively well understood as separate processes.
The collective behavior of the Coulomb collisions is usually studied in the approximation of two-particle
interactions which leads to the Boltzmann integral equation. This equation may be further approximated
by considering only small-angle scattering events; this leads to a
differential Fokker-Planck equation which considerably simplifies the
modeling of Coulomb collisions. However, it is important to keep in mind that the Fokker-Planck approximation
is not generally applicable (with respect to the Boltzmann integral approximation) even for Maxwellian particle
 velocity distribution functions  \citep{shou87}.
Coulomb collisions have a kinetic nature. The scattering cross section is strongly dependent on the relative
particle velocity and
a collisionality in different regions of the velocity distribution function varies significantly.

Kinetic instabilities in plasmas are typically studied in homogeneous collisionless plasmas.
The linear properties of these instabilities are generally described by a complicated system
which needs to be treated numerically except in
some limiting cases. Kinetic instabilities tend to reduce the source of free energy  but 
 their nonlinear properties are not fully understood.
Assuming a superposition of weak/noninteracting random phase modes leads to a quasi-linear approximation
with a diffusion in the velocity space owing to averaged second order terms \citep{keen66,hetr12}. 
Presence of important wave-wave interactions, wave coherence or of an important population
of trapped particles, however, makes this approach
questionable. 

 Sufficiently strong Coulomb collisions invalidate the collisionless approach. 
Particle trajectories are modified by collisions so that wave-particle
interactions and particle trapping are likely importantly influenced by the strong collisions.
The collisions typically introduce a damping so that one expects that they increase instability thresholds or even
suppress instabilities. 
On the other hand, wave-particle interactions often lead to
formation of small-scale structures in the particle velocity distribution functions.
Large gradients connected with such structures would lead to an enhanced collisional diffusion in
the Fokker-Planck approach \cite[cf.,][]{scheal08b}. However, the Fokker-Planck equation is
only an approximation of an approximation of another approximation. From the more general Boltzmann integral
approach  it is not clear that strong gradients in the particle velocity distribution functions
 importantly affect the collisional interaction.

The situation becomes more complicated when there is an external driver
which generates the temperature anisotropy. The system behavior
depends on relative time scales of the three process. However,
both Coulomb collisions and instabilities are kinetic so that their
characteristic times are different in
different regions of the particle velocity distribution function.
For collisions one can define an average macroscopic time scale but
in the case of instabilities it is more complicated. One possible measure of the instability time
scale is the (maximum) growth rate which typically increases monotonically above threshold.
 Hybrid collisionless simulations
with external drivers (expansion, velocity shears)
indicate that as the system enters the unstable regions waves are generated. When these waves 
haver strong enough amplitude they are able to some extent counteract the driven anisotropization 
and to keep the system near the marginal stability \citep{hetr05,mattal06,kunzal14}. This evolution depends on the time scale
of the driver. In the driven simulations   
the system enters further to the unstable region and generates stronger wave activity  for a faster driver.
This process can be to some extent modeled by a bounded anisotropy model
\citep{dental94,hetr08,chanal11} where the linear prediction is used to impose 
bounds on the temperature anisotropy by introducing a strong effective isotropization frequency
in the unstable region.

The feedback of Coulomb collisions and kinetic instabilities on the large scale
driver is another open problem.
The effects of the Coulomb collisions on the macroscopic level may be expressed by
transport coefficients which are typically derived in the vicinity
of thermal equilibrium  \citep{brag65}. A presence of other kinetic effects
may render questionable these collision-dominated theoretical predictions (in the collisionless limit the transport coefficients 
of \cite{brag65} diverge).
Some collisional transport coefficients can be calculated even far from thermal equilibrium,
for example assuming drifting bi-Maxwellian velocity distribution functions
for all particle species \citep{basc81,hetr09}. For instance the
proton-proton isotropization frequency $\nu_{\mathrm{pp}}$ 
\begin{equation}
\left[\frac{\mathrm{d} (T_{\mathrm{p}\perp} - T_{\mathrm{p}\|})}{\mathrm{d}t}  \right]_{\mathrm{coll}}
= -\nu_{\mathrm{pp}} \left( T_{\mathrm{p}\perp}-T_{\mathrm{p}\|} \right)
\label{isotropization}
\end{equation}
 may be given as
\begin{equation}
\nu_{\mathrm{pp}}=
\frac{ e^4 n_\mathrm{p} \ln\Lambda}
{10\upi^{3/2}\epsilon_0^2 m_{\mathrm{p}}^{1/2} k_B^{3/2} T_{\mathrm{p}\|}^{3/2} }
\ _2F_1
\left(\begin{array}{c}2,3/2\\ 7/2\end{array},1- \frac{T_{\mathrm{p}\perp}}{T_{\mathrm{p}\|}} \right)
\label{nupp}
\end{equation}
where \!$\ _2F_1$ is the standard (Gauss) hypergeometric function. In collisionless plasmas
kinetic instabilities lead to effective transport coefficients.  In the quasi-linear approximation 
it is possible to derive some of the transport coefficients 
\citep{yose12,hellal13b} assuming the particle velocity distribution 
function is close to bi-Maxwellian. 

The behavior of a driven system with Coulomb collisions and temperature anisotropy-driven instabilities
is a complex nonlinear problem which is hard to investigate analytically so that a numerical approach
is needed. 
In this paper we investigate expansion driven proton temperature anisotropies
 in high beta, weakly collisional plasmas. 
In high beta (collisionless) plasmas there are two dominant instabilities driven by the proton temperature anisotropy.
For $T_{\mathrm{p}\perp}>T_{\mathrm{p}\|}$ it is the mirror instability \citep{hase69}. 
This instability is resonant (i.e., a substantial portion of the proton velocity distribution
function resonates with the unstable waves \cite[cf.,][]{gary93}) through the Landau resonance. This kinetic
feature is combined with fluid properties, the unstable waves are nonpropagating and
have long wavelengths near threshold. The nonlinear properties of the mirror instability
are not fully understood, as they seem to combine kinetic properties (the Landau resonance,
particle scattering/trapping) and a fluid nonlinearity \citep{calial08}.
For $T_{\mathrm{p}\perp}<T_{\mathrm{p}\|}$  the dominant growing mode is is the oblique fire hose instability
\citep{hema00}. This instability is resonant through the cyclotron resonance and
generates transient nonpropagating modes which eventually become propagating and damped \citep{hetr08}.
Other instabilities (ion cyclotron, parallel fire hose, Weibel) may also play an important
role in regulating the proton temperature anisotropy. 

This paper is organized as follows: section~\ref{heb} describes the numerical code,
section~\ref{simul} presents the simulation results for one simulation for the 
parallel proton temperature anisotropy 
 $T_{\mathrm{p}\perp}<T_{\mathrm{p}\|}$ and one for the perpendicular anisotropy  $T_{\mathrm{p}\perp}>T_{\mathrm{p}\|}$.
The simulations results are summarized and discussed in section~\ref{discussion}.

\section{Expanding box model}

\label{heb}

Here we use 
the expanding box model \citep{grapal93}
implemented to the
hybrid code developed by \cite{matt94} 
to study a response of a weakly collisional plasma to a slow expansion.
In this Collisional Hybrid Expanding Box (CHEB)  model
the expansion is described as an external force.
This model was developed in the context of the radial expansion of the
solar wind.
One assumes a solar wind with a
constant radial velocity $U$ at a radial distance $R$.
Transverse scales (with respect to the radial direction)
 of a small portion of plasma, co-moving with the solar
wind velocity, increase  with time as
$1+t/t_e$ where
$t_e=R_0/U$ is the (initial) characteristic expansion time.
The expanding box uses these co-moving coordinates,
the physical transverse scales of the simulation
box increase with time \cite[see ][for a detailed description
of the (collisionless) code]{hetr05} and the standard periodic boundary conditions are used.
Coulomb collisions in the code are modelled using the Langevin stochastic forcing
corresponding to the Fokker-Planck equation \citep{manhal97} where particle velocity distribution
functions are assumed to be drifting bi-Maxwellian ones; in this case Rosenbluth potentials
can be expressed in terms of generalized triple hypergeometric functions and
the corresponding Langevin stochastic equation can be constructed
 \cite[cf.,][]{hetr10}.

The characteristic spatial and temporal units used in the model
are the initial proton inertial length $d_{\mathrm{p}0}=c/\omega_{p\mathrm{p}0}$ and
the inverse initial proton cyclotron frequency $1/\omega_{c\mathrm{p}0}$.
Here $c$ is the speed of light, $\omega_{p\mathrm{p}0} = ({n_{\mathrm{p}0}
e^2}/{m_\mathrm{p}\epsilon_0})^{1/2}$ is the initial proton plasma
frequency, $\omega_{c\mathrm{p}0} = {eB_{0}}/{m_\mathrm{p}}$,
$B_{0}$ is the initial magnitude of the ambient magnetic
 field $\boldsymbol{B}_0$,
$n_{\mathrm{p}0}$ is the initial proton density,
$e$ and $m_\mathrm{p}$ are the proton electric
charge and mass, respectively; finally,
$\epsilon_0$ is the dielectric permittivity of
vacuum.
We use the spatial resolution
$\Delta x = \Delta y = 2 c/\omega_{p\mathrm{p}0}$, and there are initially 16.384 particles per cell
for protons.
Fields and moments are defined on a 2-D grid  $512 \times 512$.
 Protons are advances using
the Boris' scheme with a time step $\Delta t=0.05/\omega_{c\mathrm{p}0}$,
while the magnetic field $\boldsymbol{B}$
is advanced with a smaller time step $\Delta t_B = \Delta t/10$.
The collisional Langevin stochastic forcing is applied
every 10 time steps.
The initial ambient magnetic field is directed along $x$ direction,
$\boldsymbol{B}_{0}=(B_{0},0,0)$.

We analyze two simulations,
one for the
parallel proton temperature anisotropy
 $T_{\mathrm{p}\perp}<T_{\mathrm{p}\|}$ and one for the perpendicular anisotropy  $T_{\mathrm{p}\perp}>T_{\mathrm{p}\|}$.
We assume relatively high proton betas ($\sim 20$) and relatively high Coulomb collisional rates
(with respect to the expansion time). These parameters are quite larger than those typically observed in
the solar wind as here we are more interested in high-beta astrophysical plasmas (such as
the intercluster plasmas where, 
however, even much larger plasma betas are expected).

\section{Simulation results}
\label{simul}

\subsection{Fire hose instabilities}

In the first (fire hose), 2-D CHEB simulation we initialize protons as an
isotropic Maxwellian velocity distribution function with $\beta_{\mathrm{p}}=20$.
We impose  a continuous expansion
 in the $y$ and $z$ directions (transverse with respect to the ambient magnetic field) with
the (initial) characteristic expansion time $t_e=10^4 /\omega_{c\mathrm{p}0}$.
The proton density  and the  magnitude
of the ambient magnetic field decrease as $(1+{t}/{t_e})^{-2}$ due to the expansion. Such a decrease would
lead in a collisionless plasma to the double adiabatic evolution
when no wave activity or heat fluxes are present \citep{chewal56}:
\begin{equation}
\left(\frac{\mathrm{d} T_{\mathrm{p}\perp}}{\mathrm{d}t}\right)_\mathrm{CGL}=  \frac{T_{\mathrm{p}\perp}}{B} \frac{\mathrm{d}B}{\mathrm{d}t} \ \
\mathrm{and} \ \
\left(\frac{\mathrm{d} T_{\mathrm{p}\|}}{\mathrm{d}t}\right)_\mathrm{CGL}=  
2\frac{T_{\mathrm{p}\|}}{n} \frac{\mathrm{d}B}{\mathrm{d}t}-2\frac{T_{\mathrm{p}\|}}{B} \frac{\mathrm{d}B}{\mathrm{d}t};
\label{CGL}
\end{equation}
in this case $T_{\mathrm{p}\perp}/T_{\mathrm{p}\|}$ would decrease with time.
For the Coulomb collision we set the initial isotropization
frequency $\nu_{\mathrm{pp}}=1.2 \times 10^{-3}\omega_{c\mathrm{p}}$. 
The expanding collisional system is expected to follow an evolution where
double-adiabatically driven development of the temperature anisotropy is counteracted by Coulomb collisions (when no wave activity or heat fluxes are present) 
\begin{equation}
\frac{\mathrm{d} T_{\mathrm{p}\perp,\|}}{\mathrm{d}t}=\left(\frac{\mathrm{d} T_{\mathrm{p}\perp,\|}}{\mathrm{d}t}\right)_\mathrm{CGL}+
\left(\frac{\mathrm{d} T_{\mathrm{p}\perp,\|}}{\mathrm{d}t}\right)_{\mathrm{coll}}
\label{theor}
\end{equation}
where
\begin{equation}
\left(\frac{\mathrm{d} T_{\mathrm{p}\perp}}{\mathrm{d}t}\right)_{\mathrm{coll}}
=-\frac{1}{2}\left(\frac{\mathrm{d}T_{\mathrm{p}\|}}{\mathrm{d}t}\right)_{\mathrm{coll}}
= -\frac{\nu_{\mathrm{pp}}}{3} \left( T_{\mathrm{p}\perp}-T_{\mathrm{p}\|} \right).
\label{col}
\end{equation}
and where $\nu_{\mathrm{pp}}$ is given by Equation~(\ref{nupp})
when the proton velocity distribution functions are close to bi-Maxwellian ones

\begin{figure}
\centerline{\includegraphics[width=10cm]{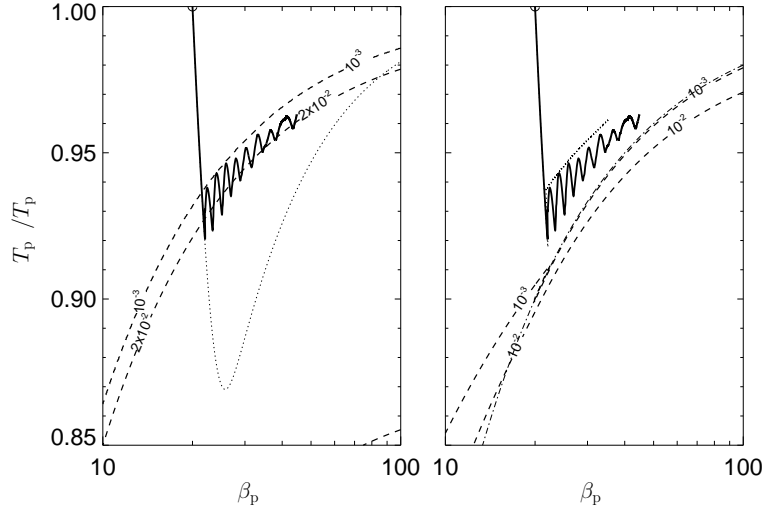}}
\caption{Solid curves denote the path in the ($\beta_{\mathrm{p}\|}$, $T_{\mathrm{p}\perp}/T_{\mathrm{p}\|}$)
space in the 2-D (fire hose) CHEB simulation.
The dashed contours show the linear prediction,
the maximum linear growth rate (in units of $\omega_{c\mathrm{p}}$) as a function of $\beta_{\mathrm{p}\|}$ and $T_{\mathrm{p}\perp}/T_{\mathrm{p}\|}$ 
for (left) the oblique fire hose and (right) for the parallel fire hose.
The dotted line on the left panel shows the theoretical collisional evolution, Eq.~(\ref{theor})
 whereas  on the right panel it shows the theoretical prediction for a simple, one-mode
quasi-linear model.
The dash-dotted line displays the fluid fire hose threshold $C_\mathrm{F}=0$.
\label{colfh}
}
\end{figure}

The evolution in the expanding system is shown in Figure~\ref{colfh}
in the ($\beta_{\mathrm{p}\|}$, $T_{\mathrm{p}\perp}/T_{\mathrm{p}\|}$)
space.
The solid curves denote the path 
in the 2-D (fire hose) CHEB simulation where
$\beta_{\mathrm{p}\|}$ and $T_{\mathrm{p}\perp}/T_{\mathrm{p}\|}$ are averaged quantities over the simulation box
calculated with respect to the ambient magnetic field.
The dashed contours show the linear (collisionless) prediction,
the maximum linear growth rate (in units of $\omega_{c\mathrm{p}}$) as a function of $\beta_{\mathrm{p}\|}$ and $T_{\mathrm{p}\perp}/T_{\mathrm{p}\|}$
for the oblique fire hose (left panel) and for the parallel fire hose (right panel).
The dotted line on the left panel shows the theoretical collisional evolution (obtained from Eq.~(\ref{theor})) whereas the
dash-dotted line displays the fluid fire hose threshold $C_\mathrm{F}=0$ where
\begin{equation}
C_\mathrm{F}=\beta_\|-\beta_\perp-2.
\label{fluid}
\end{equation}
Figure~\ref{colfh} shows that initially the system follows the theoretical
prediction of Eq.~(\ref{theor}); $T_{\mathrm{p}\perp}/T_{\mathrm{p}\|}$ decreases as the Coulomb collisions are not strong enough
to keep the protons isotropic. The initial behavior of the systerm is almost
double adiabatic as the collisional isotropization is
proportional to $|T_{\mathrm{p}\perp}-T_{\mathrm{p}\|}|$.
As the temperature anisotropy grows the system departs from the double adiabatic expectation, and, eventually, the
collisional term would dominate and the system would asymptotically reach the isotropy. Note that
the expansion driving decreases with time (as the expansion time $R/U$ increases).
In the simulation, however, the system 
becomes unstable with respect to
the oblique fire hose before the collisions become dominant and the proton
temperature anisotropy is strongly reduced. After this  $T_{\mathrm{p}\perp}/T_{\mathrm{p}\|}$ decreases again
and this evolution periodically repeats itself. In the 2-D (fire hose) CHEB 
the system is linearly stable with respect to the proton parallel fire hose which has a higher threshold
close to the fluid one.\footnote{ 
The oblique fire hose has a lower threshold compared to
that of the parallel fire hose at least up to $\beta_{\mathrm{p}\|}=10^4$; however, the two
thresholds approach each other in the
$(\beta_{\mathrm{p}\|},T_{\mathrm{p}\perp}/T_{\mathrm{p}\|})$ plane as $\beta_{\mathrm{p}\|}$ increases.}

To test the relevance of the (collisionless) quasi-linear approximation for
the description of the oblique fire hose we have developed a simple, one-mode quasi-linear model
for the instability as a more physical version of a bounded anisotropy model. In this model only one mode is considered and is assumed to have
the growth-rate and quasi-linear diffusion properties of the most unstable mode. In
the stable region the amplitude of this mode is assumed to be constant (the mode starts with a small, dynamically
negligible amplitude) and no diffusion is assumed whereas in
the (collisionless) unstable region this mode grows following the (collisionless) prediction based
on bi-Maxwellian VDFs $\mathrm{d}\delta B/\mathrm{d} t= \gamma_{max}\delta B$ (changes
in the position of the most unstable mode in the wave vector space are ignored);
the quasi-linear heating rates 
 $\left(\mathrm{d} T_{\mathrm{p}\perp,\|}/\mathrm{d}t\right)_\mathrm{QL}$
are calculated for the most unstable mode also assuming bi-Maxwellian VDFs \citep{hellal13b}.
The result of this model is shown on Figure~\ref{colfh} (right panel, dotted line).
The simple model captures relatively well the initial excursion of the system to the unstable
region and after the saturation, it
predicts a marginal stability path with respect to the oblique fire hose instability.

\begin{figure}
\centerline{\includegraphics[width=10cm]{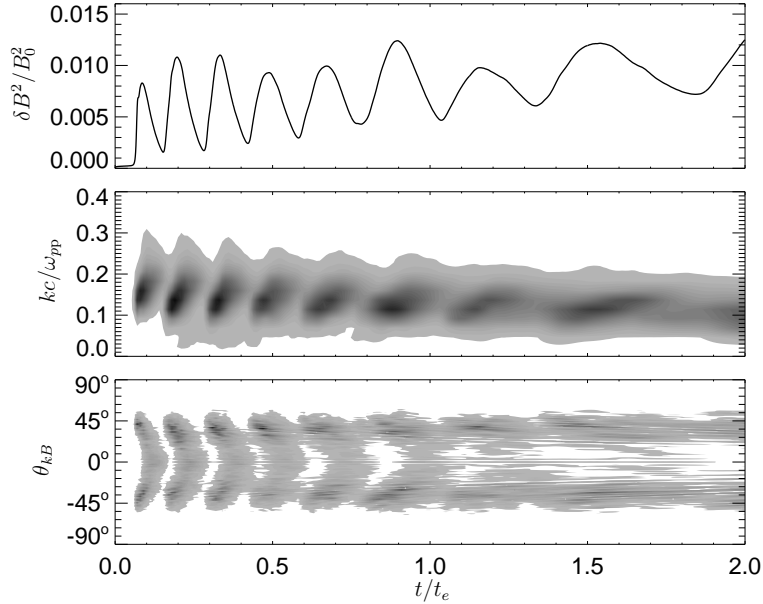}}
\caption{Evolution of the magnetic fluctuations in the 2-D (fire hose) CHEB simulation:
(top) Fluctuating magnetic field $\delta B^2/B_0^2$ as a function of time.
Gray scale plots of
the fluctuating magnetic field $\delta B$ as a function of time and $k$
(middle panel) and as a function of time and $\theta_{kB}$
(bottom panel).
\label{dbfh}
}
\end{figure}

The evolution of the wave spectrum in the simulation is shown in Figure~\ref{dbfh}.
Figure~\ref{dbfh} (top panel) displays the fluctuating magnetic field $\delta B^2/B_0^2$ as a function of time.
The middle and bottom panels show gray scale plots of
the fluctuating magnetic field $\delta B$ as a function of time and the wave vector $k$
and as a function of time and the propagation angle $\theta_{kB}$, respectively.
Figure~\ref{dbfh} shows that in the weakly collisional plasma the oblique fire hose has 
an evolution similar to that of the collisionless case \citep{hetr08}. The oblique fire hose generates
waves at oblique angles with respect to the ambient magnetic field. The spectrum evolves towards
less oblique angles and most of the fluctuating energy is damped. The system with a low wave activity
evolves roughly following the weakly collisional prediction till enough oblique fire hose wave energy 
is again generated and reduces the proton temperature anisotropy.  This behavior continues in a semiperiodic manner
with a period which seems to increase with time. On average the magnetic fluctuating energy
increases with time, after each period there is more remaining fluctuating magnetic energy.

The linear and quasi-linear analyses \citep{hema00,hema01} indicate that this behavior may be understood
in the quasi-linear framework where a dispersion change is included. The oblique fire hose instability
generates non-propagating modes which only exist for sufficiently anisotropic protons. As the generated
waves scatter protons and reduce the proton anisotropy the non-propagating modes no longer exist and
the corresponding wave energy must be transformed mainly to standard, ion cyclotron waves. These waves are relatively strongly
damped by protons through the cyclotron resonance.
In the collisionless plasmas this leads to a deformation of the proton VDFs in the resonant
regions \citep{mattal06,hetr08}. This deformation is not seen in the present simulation as the Coulomb
collisions are relatively strong (and the resonant interaction in the high beta plasma may be less
obvious or possibly less effective, cf., Matteini \textit{et al.} 2006). The proton VDFs (averaged over the box) remains relatively close
to bi-Maxwellian ones during the simulation, so that the two parallel and perpendicular temperatures
characterize very well the proton VDFs. This justifies the usage of the bi-Maxwellian Langevin model
and, similarly, this may justify the application of the quasi-linear heating rates
based on the bi-Maxwellian VDFs for the macroscopic description of the oblique fire hose instability;
the one-mode model used here is, however, too simple to capture the evolution of the system since 
it does not include the branch change from nonpropagating
to propagating modes.

\subsection{Mirror and ion cyclotron instabilities}

In the second (mirror), 2-D CHEB simulation we initialize protons with
isotropic bi-Maxwellian velocity distribution function with $\beta_{\mathrm{p}}=30$.
In this case we impose  a continuous expansion
 in the $x$ and $y$ directions with
the (initial) characteristic expansion time $t_e=10^4 /\omega_{c\mathrm{p}0}$
(the radial direction in the model of expanding solar wind would be in the $z$ direction,
i.e., perpendicular to $\boldsymbol{B}_0$ which is aligned with the $x$ axis).
The expansion leads
to a decrease of the density as $(1+t/t_e)^{-2}$ whereas the magnitude
of the magnetic field decreases as  $(1+{t}/{t_e})^{-1}$.
In this case the double adiabatic evolution would lead to increasing $T_{\mathrm{p}\perp}/T_{\mathrm{p}\|}$.
The collisional isotropization
frequency $\nu_{\mathrm{pp}}$ is chosen be $6.5 \times 10^{-4}\omega_{c\mathrm{p}}$.

\begin{figure}
\centerline{\includegraphics[width=10cm]{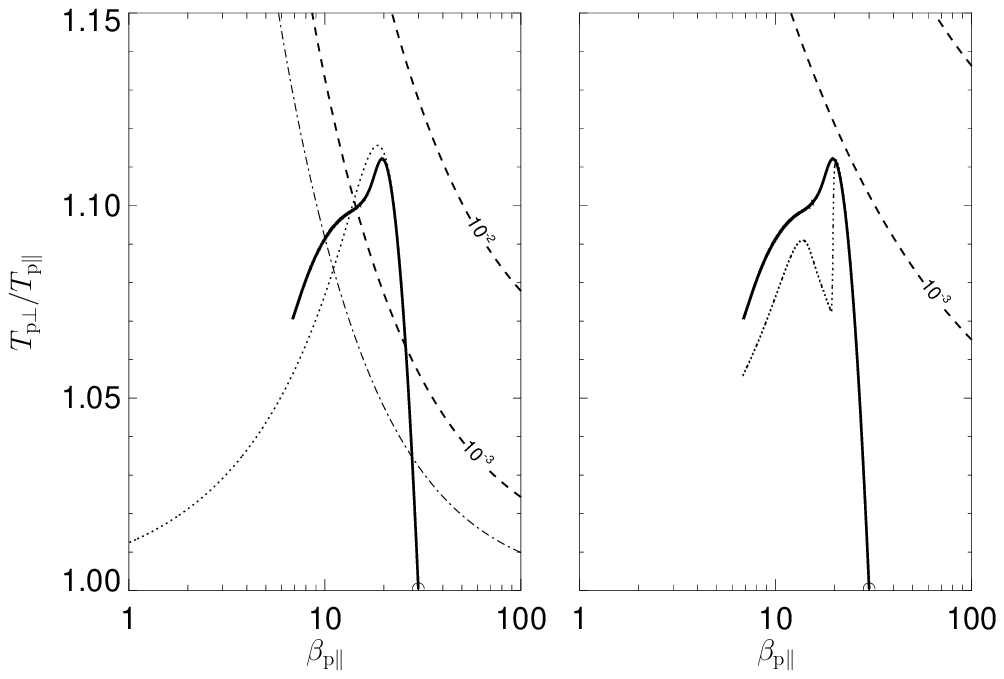}}
\caption{
Solid curves denote the path in the ($\beta_{\mathrm{p}\|}$, $T_{\mathrm{p}\perp}/T_{\mathrm{p}\|}$)
space in the 2-D (mirror) CHEB simulation.
 The dashed contours show the linear prediction,
the maximum linear growth rate (in units of $\omega_{c\mathrm{p}}$) as a function of $\beta_{\mathrm{p}\|}$ and $T_{\mathrm{p}\perp}/T_{\mathrm{p}\|}$
for (left) mirror instability and (right) for the ion cyclotron instability.
The dotted line on the left panel shows the theoretical collisional evolution, Eq.~(\ref{theor}),
 whereas  on the right panel it shows the theoretical prediction for a simple, one-mode
quasi-linear model. The dash-dotted line displays the mirror threshold $C_\mathrm{M}=0$.
\label{colmir}
}
\end{figure}

Figure~\ref{colmir} shows the evolution of the system in the  2-D (mirror) CHEB simulation.
Solid curves denote the path in the ($\beta_{\mathrm{p}\|}$, $T_{\mathrm{p}\perp}/T_{\mathrm{p}\|}$) on
both the panels  ($\beta_{\mathrm{p}\|}$ are $T_{\mathrm{p}\perp}/T_{\mathrm{p}\|}$ averaged over the box as in Figure~\ref{colfh}). The dashed contours show the linear prediction,
the maximum linear growth rate (in units of $\omega_{c\mathrm{p}}$) as a function of $\beta_{\mathrm{p}\|}$ and $T_{\mathrm{p}\perp}/T_{\mathrm{p}\|}$
for the mirror instability (left panel) and for the ion (proton) cyclotron instability (right panel).
The dotted line on the left panel shows the theoretical collisional evolution (given by Eq.~(\ref{theor})) whereas the
dash-dotted line displays the mirror threshold $C_\mathrm{M}=0$ \cite[cf.,][]{chanal58,hell07}
 where 
\begin{equation}
C_\mathrm{M}=\beta_{\mathrm{p}\perp} (T_{\mathrm{p}\perp}/T_{\mathrm{p}\|}-1)-1-\frac{1}{2} 
\frac{(T_{\mathrm{p}\perp}/T_{\mathrm{p}\|}-1)^2}{\beta_\mathrm{e}^{-1}+\beta_{\mathrm{p}\|}^{-1}}.
\label{mirrorth}
\end{equation}

Figure~\ref{colmir} shows that the system initially follows the theoretical collisional prediction of
Eq.~(\ref{theor}). Again, the initial behavior of the systerm is almost
double adiabatic but
as the temperature anisotropy increases the system departs from the double adiabatic expectation, and, eventually, the
collisional term would dominate and the system would asymptotically reach the isotropy.
 In the region unstable with respect to the mirror instability
(in the corresponding collisionless plasma) the temperature anisotropy is reduced faster compared to
the collisional prediction due to the generation of (mostly) mirror modes.  
This reduction continues in a manner qualitatively similar
to the theoretical prediction likely mainly due to the Coulomb collisions.
The dotted line on the right panel of Figure~\ref{colmir} shows a prediction of the simple
one-mode quasi-linear model. It captures some of the properties of the simulated system,
namely the fast
reduction of the anisotropy. This model also exhibits an evolution following a marginal stability path
with respect to the mirror instability.

\begin{figure}
\centerline{\includegraphics[width=10cm]{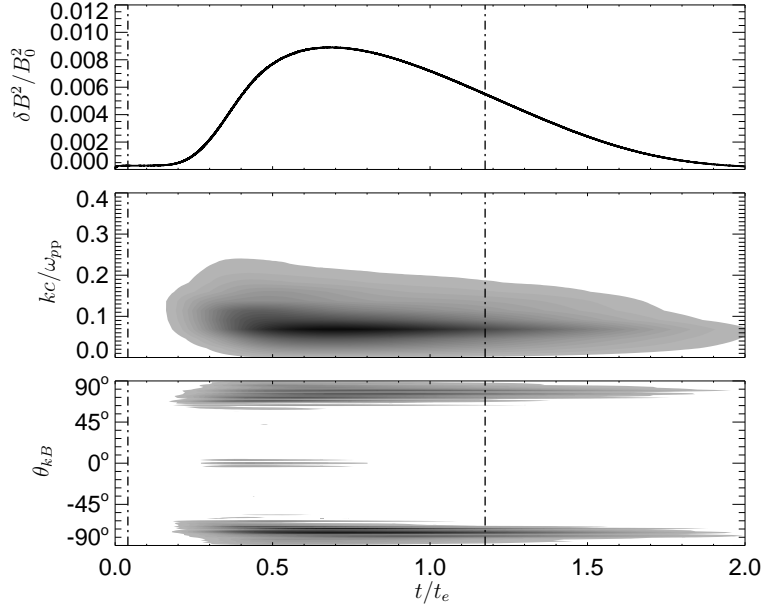}}
\caption{Evolution of the magnetic fluctuations in the 2-D mirror CHEB simulation:
(top) Fluctuating magnetic field $\delta B^2/B_0^2$ as a function of time.
Gray scale plots of
the fluctuating magnetic field $\delta B$ as a function of time and  $k$
(middle panel) and as a function of time and $\theta_{kB}$
(bottom panel). The dash-dotted lines display the time when the system
crosses the mirror threshold $C_\mathrm{M}=0$.
\label{dbmir}
}
\end{figure}

The temporal evolution of the wave spectrum is shown in Figure~\ref{dbmir}:
The top panel displays the fluctuating magnetic field $\delta B^2/B_0^2$,
the middle and bottom panels show gray scale plots of
the fluctuating magnetic field $\delta B$ as a function of time and the wave vector $k$
and as a function of time and the propagation angle $\theta_{kB}$, respectively.
 The dash-dotted lines display the times when the system
crosses the mirror threshold $C_\mathrm{M}=0$ (see Figure~\ref{colmir}).
Figure~\ref{dbmir} shows that the fluctuating magnetic energy is generated
later on in the simulation when the system is in the unstable region and this energy is then damped;
at the end the system is in a region stable with respect to the mirror instability.
The generated modes are at mostly strongly oblique angles which correspond to the mirror modes. There also weak
signatures of quasi-parallel waves which correspond to the proton cyclotron modes.

The mirror modes remain in the system for a relatively long time even in the region linearly stable with
respect to the mirror instability. This is likely related with the nonlinear properties (bistability)
of the mirror modes. The mirror instability tends to create coherent structures
either in the form of magnetic enhancements (humps) or in the form
of magnetic depressions (holes). The nonlinear model of \cite{kuznal07,kuznal07b}
for nonlinear development of the mirror instability,
 near threshold  based on a reductive
perturbative expansion of Vlasov-Maxwell equations
predicts formation of magnetic holes; however, numerical
simulations exhibit rather formation of magnetic humps.
This effect is likely connected with the resonant properties of the mirror
instability. While the mirror modes near threshold appear at large scales,
their  driving mechanism is 
the Landau resonance between the nonpropagating mirror modes and protons
with small parallel velocities. Small changes of proton velocity distribution 
function in the resonant regions strongly influences the nonlinear term
derived from the perturbative expansion. This may lead
to the change of sign of the nonlinearity and to the formation of magnetic humps
instead of holes \citep{hellal09}.

\begin{figure}
\centerline{\includegraphics[width=8cm]{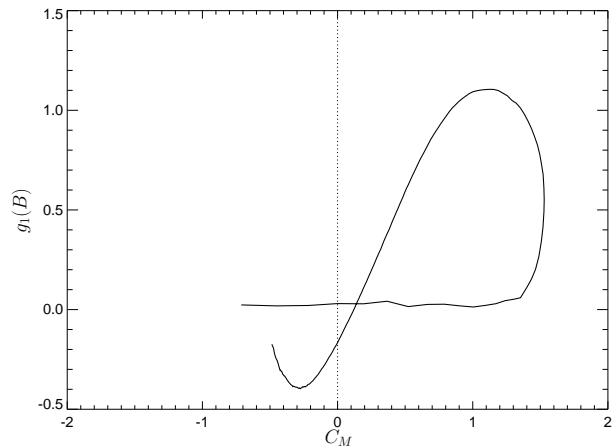}}
\caption{Evolution in the 2-D mirror CHEB code: Skewness of the amplitude
of the magnetic field $g_1(B)$ as a function of the mirror linear criterion
$C_\mathrm{M}$.
\label{skew}
}
\end{figure}

In the present CHEB simulation we observe initially a formation of magnetic humps
in the unstable region. These humps 
are transformed to magnetic holes as the system becomes linearly stable
with respect to the mirror instability. This can be seen in Figure~\ref{skew}
which displays  the (sample estimate of) skewness of the amplitude
of the magnetic field $g_1(B)$ (calculated at a given time over the 512x512 grid points) as a function of the mirror linear criterion
$C_\mathrm{M}$.
The skewness of  a variable $x$ with $N$ points is defined as 
\begin{equation*}
g_1(x)= \frac{\sqrt {N(N-1)}}{N-2}   \frac{\langle (x-\langle x\rangle)^3\rangle}{ \langle (x-\langle x\rangle)^2\rangle^{3/2} }
\end{equation*}
where $\langle x \rangle=\sum_{i=1}^{N}x_i/N$ denotes averaging.
 The system starts in the stable region with a zero skewness.
As the system moves to the unstable regions the mirror modes
are generated and the skewness becomes positive which is a sign of magnetic
humps. As the Coulomb collisions drives the system to the stable region
the skewness becomes negative indicating magnetic holes.
Similar evolution is also observed in collisionless hybrid expanding box simulations 
and there are signatures of a comparable trend in the terrestrial magnetosheath
\citep{traval07b,genoal09}.

The proton velocity distribution function
remains quite close to a bi-Maxwellian during the 
 whole simulation as the Coulomb collisions are relatively strong.
However, some weak signatures of flattening of the proton velocity distribution function in
the resonant region are also observed (this is compatible with the presence of magnetic
humps in the unstable region). Such a flattening is expected from the quasilinear predictions
\citep{calial08}. The quasilinear approximation is, however, questionable
 when coherent structures appear; on the other hand, the proton trapping
has a similar effect on the velocity distribution function \citep{pojo13}.

\section{Discussion}
\label{discussion}

In this paper we study evolution of expansion-driven temperature
anisotropy in a weakly collisional, relatively high beta plasma using the CHEB model.
We performed two 2D CHEB simulations, one for the fire hose instabilities and one
for the mirror and ion cyclotron instabilities.
The parameters for the two simulations were chosen as a compromise to have feasible runs in
a kinetic particle-in-cell hybrid code; 
large betas lead to high numerical noise so that a large number of particles per cell is
needed to have a good separation between the noise and waves generated by the instabilities. 
The inverse collisional frequency was chosen to be relatively large (but comparable) with respect to the expansion
characteristic time. For the chosen parameters the expansion generates proton temperature anisotropies
sufficient to destabilize the system despite the presence of relatively strong Coulomb collisions.

In the first CHEB simulation only the linearly dominant oblique fire hose is observed. The wave activity
generated by this instability efficiently scatters protons and effectively reduces the proton temperature
anisotropy (leading to an effective isotropization frequency much faster the collisional one). 
This instability has a self-stabilizing properties, most of the generated wave energy   
is reabsorbed by protons. The evolution exhibits an oscillation between unstable and marginally stable
states with generation and absorption of the wave energy.
In the second CHEB simulation we observed mainly the mirror mode activity. The mirror instability
generates coherent structures in the form of magnetic humps which scatter protons. As the Coulomb
collisions reduce the temperature anisotropy the system becomes stable with respect to the mirror
instability and the magnetic humps transform into magnetic holes which survive relatively long
time in the stable region. We developed a simple, one-mode model based on
quasi-linear heating rates for bi-Maxwellian velocity distribution functions. This model
exhibit some basic properties of the two instabilities; a more general quasi-linear model may be useful
for modeling of the macroscopic effects of the oblique fire hose instability and possibly
of the mirror instability to some extent. More theoretical and numerical work is needed. 

We used the expansion as a driver for the temperature anisotropy.
In many astrophysical situations such as in the intracluster medium it is expected that
a large scale (turbulent) motion leads to velocity shears which
drive the temperature anisotropy. We expect that a similar evolution we observed
in the case of expansion
takes place also in the case of velocity shears. Indeed, recent (collisionless) hybrid shearing box
simulations \citep{kunzal14} show similar properties.
In the cases when the velocity shear generates $T_{\mathrm{p}\perp}<T_{\mathrm{p}\|}$
 \cite{kunzal14} observe oblique fire hose instability which efficiently reduces
the temperature anisotropy. However, \cite{kunzal14} don't observe oscillatory
behavior comparable to our results. This is possibly due to the fact that
 \cite{kunzal14} stop their simulations shortly after the fluctuating magnetic
energy reaches the maximum. Moreover, our CHEB results indicate that the decaying time increases with
the proton beta (and additional standard hybrid simulations of the oblique
fire hose instability up to $\beta_{\mathrm{p}\|}=100$ confirm this trend) 
and \cite{kunzal14} investigated cases with $\beta_{\mathrm{p}\|}\sim 200$.
In the cases when the velocity shear generates $T_{\mathrm{p}\perp}>T_{\mathrm{p}\|}$
\cite{kunzal14} observe mirror mode structures which trap and scatter protons.
 This is similar to our results. We note, however, that in our case relatively strong Coulomb collisions are
present and scatter protons which likely reduces proton trapping. This may possibly
contribute to the quasi-linear diffusion.\footnote{Particle trapping and quasi-linear diffusion 
are two connected phenomena. A quasi-linear-like diffusion appears
when the trapped regions for different noncoherent modes sufficiently overlaps \citep{chir79}.}

In our weakly collisional simulations the instabilities have nonlinear properties
relatively similar to those in collisionless plasmas. In the case of
oblique fire hose instability the nonlinear evolution which may be understood in
the quasi-linear framework including a branch change. In this
case the macroscopic properties may be partly described by a combination of
collisional and quasi-linear terms. The case of the mirror instability
is more complicated. The mirror instabilities generates nonlinear structure
which evolve in time and survive relatively long into the stable region
(introducing a memory in the system). The mirror structures reduce the anisotropy
by scattering protons which can be only partially described as a quasi-linear
diffusion. In our CHEB simulations (as well as in simulations using the shearing
box model) the external driver is modeled as an independent force. Therefore, these
models cannot be used to study the feedback of temperature-anisotropy-driven instabilities
on the driver. For instance, general properties of the viscosity tensor remain an
open problem \cite[cf.,][]{mosc14}.

\section*{Acknowledgements}

Authors acknowledge the grant P209/12/2023
 of the Grant Agency of the Czech Republic.
The research leading to these results has received funding from the
    European Commission's Seventh Framework Programme (FP7) under
    the grant agreement SHOCK (project number 284515, project-shock.eu).
This work was also supported by the projects RVO:67985815 and RVO:68378289.

\end{document}